\documentclass[a4paper,aip,pop,twocolumn,superscriptaddress,reprint]{revtex4-1}
\usepackage{amssymb}
\usepackage{amsmath}
\usepackage{amssymb}
\usepackage{amsfonts}
\usepackage{amsthm}
\usepackage{latexsym}
\usepackage{hyperref}
\usepackage{cleveref}
\usepackage{graphicx}
\usepackage{enumerate}
\usepackage{bm}
\usepackage[utf8]{inputenc}

\hypersetup{
    colorlinks,
    citecolor=blue,
    linkcolor=blue,urlcolor=blue
}

\begin{document}

\title{Laser-plasma interactions with a Fourier-Bessel Particle-in-Cell method}

\author{Igor A. Andriyash}
\email{igor.andriyash@gmail.com}
\affiliation{Synchrotron Soleil, L’Orme des Merisiers, Saint Aubin, 91192 Gif-sur-Yvette, France}
\affiliation{LOA, ENSTA ParisTech, CNRS, Ecole polytechnique, Université Paris-Saclay, 828 bd des Maréchaux, 91762 Palaiseau cedex France}
\author{Remi Lehe}%
\affiliation{Lawrence Berkeley National Laboratory, Berkeley, CA 94720, USA}
\author{Agustin Lifschitz}
\affiliation{LOA, ENSTA ParisTech, CNRS, Ecole polytechnique, Université Paris-Saclay, 828 bd des Maréchaux, 91762 Palaiseau cedex France}%

\date{\today}

\begin{abstract}
A new spectral particle-in-cell (PIC) method for plasma modeling is presented and discussed. In the proposed scheme, the Fourier-Bessel transform is used to translate the Maxwell equations to the quasi-cylindrical spectral domain. In this domain, the equations are solved analytically in time, and the spatial derivatives are approximated with high accuracy. In contrast to the finite-difference time domain (FDTD) methods that are commonly used in PIC, the developed method does not produce numerical dispersion, and does not involve grid staggering for the electric and magnetic fields. These features are especially valuable in modeling the wakefield acceleration of particles in plasmas. The proposed algorithm is implemented in the code PLARES-PIC, and the test simulations of laser plasma interactions are compared to the ones done with the quasi-cylindrical FDTD PIC code CALDER-CIRC.

\end{abstract}

\maketitle

\section{\label{intro} Introduction}

In the last few decades, numerical simulation has become an indispensable tool for plasma physics in a wide range of problems from astro- and atmosphere physics, to laser plasma interactions. Often for such studies one needs to model plasma kinetics in presence of the strong electromagnetic fields, which produce relativistic motion of the charges in plasma. Such systems can be described by using the particle methods \cite{hockney:1988}, where plasma is presented by the {\it macro-particles}, which have the same charge-to-mass ratios as plasma particle species, but represent large quantities of the real ones. One example here is the particle-in-cell (PIC)  method, which is widely used for the plasma modeling \cite{langdon:2004}. In PIC, the electromagnetic fields are calculated at the nodes of a spatial grid, and each macro-particle interacts with only a few such nodes in its vicinity. 

The majority of the PIC codes advance the electromagnetic fields via the finite-difference time domain (FDTD) methods \cite{taflove:2005}.  In this approach, the finite differences approximate all derivatives in the Maxwell equations, and the discrete steps are used to advance the fields-particles system in time. The FDTD scheme stability is defined by the resolutions, $\Delta t$ and $\Delta\mathbf{r}$, which define the precisions of the temporal and spatial derivatives of the electric and magnetic fields. In practice, the velocities of the traveling waves in FDTD happen to depend on how well their propagation is resolved in space and time, and this phenomenon is known as  the \emph{numerical dispersion}. In the cases, where electromagnetic waves co-propagate with the relativistic particles, such dispersion may result to the unphysical wave-particle interactions, which should be treated explicitly \cite{lehe:PRSTAB2013,nuter:EPJD2014,godfrey:JCP2014}. On the other hand, the derivatives in the Maxwell equations are coupled, and for the better accuracy, the fields $\mathbf{E}$ and $\mathbf{B}$ in FDTD are usually staggered in space and in time. In this case, the terms $\mathbf{E}_p$ and $\bm{\beta}_p\times\mathbf{B}_p$ of the Lorentz force, with which wave acts on the particle, may compensate each other with a reminder $\mathbf{F}_p\sim e\mathbf{E}_p/\gamma_p^2$, where ${\gamma_p=(1-\beta_p^2)^{-1/2}}$ is the particle's Lorentz factor. For a relativistic particle, the force $\mathbf{F}_p$ can be very small, and for the correct projection of the  staggered fields a very fine temporal and spatial resolutions may be required \cite{lehe:PRSTAB2014}.

The mentioned issues become especially important in modeling the wakefield acceleration of particles (WFA). In WFA, the strong plasma waves transfer the energy from a high-power laser or particle \emph{driver} beam, to the trailing particles, eventually injected into these waves \cite{esarey:RMP2009,Litos:Nat2014,Corde:Nat2015}. These accelerators are being actively explored for the last two decades in pursuit for the compact sources of  high intensity beams of energetic particles. The three-dimensional FDTD PIC modeling is commonly used for the detailed theoretical studies of WFA physics, and to interpret the experiments. Besides the WFA modeling, such simulations are used to study the accompanying phenomena, e.g. generation of X-rays by the accelerated charges \cite{corde:RevModPhys2013}, collimation of the particle beams \cite{lehe:PRSTAB2014} etc. 

One alternative to the FDTD approach is the pseudo-spectral time domain (PSTD) method \cite{haber:PROC1973}. In PSTD, at each time step the equations are translated from the real space to a spectral (e.g. Fourier) domain, where the derivatives are presented by the linear coefficients. This approach provides high precision in calculation of the spatial derivatives, and, in some cases, allows to run simulations with the much lower spatial resolutions, than the finite differences methods \cite{LiuQH:MOP1997}. Moreover, for a wide class of equations, their spectral counterparts can be integrated in time analytically, so that the accuracy of the fields dynamics modeling does not directly depend on the simulations temporal resolution \cite{haber:PROC1973}. This is known as the pseudo-spectral analytical time domain (PSATD) method, and it was shown to produce no numerical dispersion of electromagnetic fields associated with the temporal and spatial resolutions in PIC \cite{Vay:JCP2013}. 

Here we develop the PSATD PIC considering that the model geometry has a certain level of cylindrical symmetry, which allows to replace the three-dimensional Cartesian geometry with a series of cylindrical models with different symmetries, e.g. Fourier series over the azimuthal angle. This is known as the \emph{quasi-cylindrical} geometry, and previously it was proposed for the FDTD PIC simulations in \cite{lifschitz:JCP2009}. Allowing to model the laser-plasma interactions with only few angular modes, the quasi-cylindrical FDTD approach has become popular for modeling WFA-type problems  \cite{davidson:JCP2015},  and the combinations of such method with the Fourier methods were considered \cite{YuP:arXiv2015}.  For the fully spectral quasi-cylindrical PSATD, we consider the decomposition into the \emph{cylindrical harmonics}, where the angular Fourier decomposition is naturally extended by the Fourier-Bessel transform of the azimuthal modes. Previously, the quasi-cylindrical spectral modeling was considered for a variety of problems \cite{guizar:JOCA2004,veysman:POP2006,andriyash:JCP2014}, and in this work we apply it for PIC simulations. One similar method was recently developed in \cite{lehe:arxiv2015}, in parallel with the present one. In \cite{lehe:arxiv2015} the different mathematical formulation is used, which has led to a different numerical scheme. 

The physical and mathematical models are presented and discussed in \cref{sec1}. In \cref{sec2}, we briefly discuss the implementation of the scheme, and demonstrate a few examples of laser plasma simulations, which are compared to the ones provided by the quasi-cylindrical FDTD PIC. The conclusions are given in \cref{sec3}, and are followed by a brief review of the spectral transforms (\cref{appendix1}), and their properties (\cref{appendix2}).

\section{\label{sec1} Physical and mathematical models}

Two main ingredients of a numerical model of plasma electrodynamics are the integrator of the electromagnetic equations called \emph{Maxwell solver}, and the \emph{particle pusher}, which integrates the motion equations for the charged plasma particles. In  our Maxwell solver, we employ the Fourier-Bessel transform for the spatial distributions of electromagnetic fields, and the charge densities and currents. The description of the transform is provided in \cref{appendix1}, and in the following text we denote the spectral images with the hat-like accent, $\hat{f}$. The spatial dependencies are considered in the cylindrical coordinates $\mathbf{r}=(x,r,\theta)$, while the vectorial components are chosen to be Cartesian, $\mathbf{A}=(A_x,A_y,A_z)$, so they are naturally well-defined on the axis, $r=0$ (in contrast to the cylindrical components). 

The components of the Fourier-Bessel series are the cylindrical harmonics, which are the eigenfunctions of the Laplace operator, ${\widehat{\nabla^2 f}=-\omega^2\widehat{f}}$. On the other hand, the first-order operators involve coupling of the azimuthal modes, and do not transform directly into the linear coefficients (see \cref{appendix2}). Therefore, to take advantage of these properties, we derive the mathematical model, where the Laplace operator acts on the functions updated in time (i.e. are acted on by $\partial_t$), and the operators $\bm{\nabla}$, $\bm{\nabla}\cdot$ and $\bm{\nabla}\times$ act only on the terms, which are re-calculated at each iteration.

For convenience, in the following discussion we use the dimensionless units -- the field components are normalized as $\bm{\varepsilon} = e\mathbf{E}/(mc^2k_0)$ and $\mathbf{b} = e\mathbf{B}/(m_e c^2k_0)$, where $m_e$ and $e$ are the mass of the electron and the elementary charge, $k_0=2\pi/\lambda_0$ and $\omega_0 = k_0 c$ are the wavenumber and the frequency, which correspond to a scale unity $\lambda_0$ (e.g. laser or plasma wavelength). The mass, charge, velocity, coordinates of the particles are in the units of $m_e$, $e$, $c$, $k_0^{-1}$ respectively, and time is normalized to $\omega_0^{-1}$. The particles density $n_p$ in this notations is normalized to the critical plasma density for the unity wavelength, $n_c=m_e\omega_0^2/4\pi e^2$.

\subsection{\label{sec1:1} Field equations in the spectral space}

To construct the spectral Maxwell solver, we start from the standard equations for the electric and magnetic fields (cf \cite{landau:FieldTheor1975}):
\begin{align}\label{maxwell0}
\partial_t \bm{\varepsilon} = \bm{\nabla} \times\mathbf{b}-\mathbf{j} \,,\quad &\partial_t\mathbf{b}  =  -\bm{\nabla} \times \bm{\varepsilon} \,,\\
\bm{\nabla}\cdot \bm{\varepsilon} = n\,,\quad &\bm{\nabla}\cdot \bm{b} = 0\,,\nonumber
\end{align}
where $n$ and $\mathbf{j}$ are the normalized charge density and current.

As mentioned before, in the Fourier-Bessel space the first-order operators couple the angular Fourier modes, and are to be excluded from the integration. For this, we replace the magnetic field with its rotation vector, ${\mathbf{g}=\bm{\nabla}\times \mathbf{b}}$, which in contrast to the pseudo-vector $\mathbf{b}$ is a polar vector, and it behaves similarly to the currents (in magnetostatic problems ${\mathbf{g} = \mathbf{j}}$). Next, we take the curl of the Faraday's law and combine it with the Poisson equation, which leads to:
\begin{subequations}\label{maxwell1}
\begin{equation}\label{maxwell1:1}
\partial_t \bm{\varepsilon}  = \mathbf{g} - \mathbf{j}\,,\quad \nabla^2 \bm{\varepsilon} = \partial_t \mathbf{g} + \bm{\nabla} n\,,
\end{equation}
and the equation for the magnetic field is:
\begin{equation}\label{maxwell1:2}
\nabla^2\mathbf{b}=-\mathbf{\nabla}\times \mathbf{g} \,.
\end{equation}
\end{subequations}

Equations (\ref{maxwell1}) are now well adapted for the integration in the spectral space, and after the Fourier-Bessel transform the first pair reads:
\begin{subequations}\label{maxwell2}
\begin{equation}\label{maxwell2:1}
\partial_t \widehat{\bm{\varepsilon}} - \widehat{\mathbf{g}} = - \widehat{\mathbf{j}}\,,\quad \partial_t \widehat{\mathbf{g}} + \omega^2 \widehat{\bm{\varepsilon}} = - \widehat{\bm{\nabla} n}\,,
\end{equation} 
where $\omega = \sqrt{k_x^2+ k_r^2}$, and $k_x$ and $k_r$ are the longitudinal and  transverse wavenumbers (for details see \cref{appendix1,appendix2}). Equations (\ref{maxwell2:1}) can be used to advance $\bm{\varepsilon}$ and $\mathbf{g}$ in time, and the magnetic field, needed for the Lorentz force calculation, follows from the static equation,
\begin{equation}\label{maxwell2:2}
 \widehat{\mathbf{b}} = \omega^{-2}\, \widehat{\bm{\nabla} \times\mathbf{g}}\,.
\end{equation} 
\end{subequations}

It is easy to see, that, while $\bm{\varepsilon}$ and $\mathbf{g}$ can be advanced in time with \cref{maxwell2:1}, the first-order derivatives appear only to communicate fields with particles, so they are recalculated at each integration step. 

\subsection{\label{sec1:2} Integration cycle of the quasi-cylindrical spectral PIC}

In the time-domain particle-in-cell methods, the system of fields and particles advances in time by the discrete steps. Let us consider one integration cycle on the time interval $(t_0,t_0+\Delta t)$, and denote the variables at $t_0$, $t_0+\Delta t/2$ and $t_0+\Delta t$ with the indices ``0'',``1/2'' and ``1'' respectively. During one cycle:
\begin{enumerate}[\bf (i)]
 \item particles velocities $\bm{\beta}_{1/2}$ are used to advance their positions from $\mathbf{r}_0$ to $\mathbf{r}_1$ ({\it leapfrog}),
 \item densities $n_1$ and currents $\mathbf{j}_{1/2}$ are deposed onto the spatial grid for each angular mode,
 \item spectral projections of $\widehat{\mathbf{j}}_{1/2}$ and $\widehat{\bm{\nabla} n}_1$ are calculated, 
 \item \cref{maxwell2:1} is used to update $\widehat{\bm{\varepsilon}}_1$ and $\widehat{\mathbf{g}}_1$, and $\widehat{\mathbf{b}}_1$ is calculated from \cref{maxwell2:2}
 \item $\mathbf{b}_1$ and $\bm{\varepsilon}_1$ are projected onto the spatial grid,
 \item Lorentz force is calculated at the particles positions, and their velocities are advanced to $\bm{\beta}_{3/2}$.
\end{enumerate}
The steps ({\bf i,ii,vi}) are common for the PIC methods, and their various implementations are widely discussed in the literature \cite{langdon:2004}. The steps ({\bf iii}) and ({\bf v}), are simply the translations between spectral and real domains, and these operations are discussed in \cref{appendix2}.

The step ({\bf iv}) corresponds to the spectral Maxwell solver, where eqs. (\ref{maxwell2:1}) are integrated in time for $\widehat{\bm{\varepsilon}}$ and $\widehat{\mathbf{g}}$. Let us assume that, following the particles dynamics, the current $\mathbf{j}_{1/2}$ remains constant during one integration cycle, and the density evolves linearly from $n_0$ to $n_1$. In this case, on the interval $(t_0,t_0+\Delta t)$ the electromagnetic equations can be written as:
\begin{align}\label{maxwell4}
&\partial_t \widehat{\bm{\varepsilon}} = \widehat{\mathbf{g}}- \widehat{\mathbf{j}}_{1/2}\,,\\
&\partial_t \widehat{\mathbf{g}} = - \omega^2 \widehat{\bm{\varepsilon}}  + \frac{t-\Delta t-t_0 }{\Delta t}\; \widehat{\bm{\nabla} n}_0 - \frac{t - t_0}{\Delta t}\;\widehat{\bm{\nabla} n}_1\,.\nonumber
\end{align}
\Cref{maxwell4} can be integrated in time analytically, which allows to obtain the fields at $t_0+\Delta t$ as:
\begin{equation}\label{maxwell5}
\widehat{\bm{\varepsilon}}_1 = \mathbf{C}_\varepsilon \cdot \mathbf{S}\,,\quad \widehat{\mathbf{g}}_1 = \mathbf{C}_g \cdot \mathbf{S}\,,
\end{equation}
where vector $\mathbf{S} = \Big(\widehat{\bm{\varepsilon}}_0\,,\; \widehat{\mathbf{g}}_0 \,,\; \widehat{\mathbf{j}}_{1/2} \,,\; \widehat{\bm{\nabla} n}_0 \,,\;\widehat{\bm{\nabla} n}_1\Big)$ contains the known variables, and the integration coefficients are:
\begin{widetext}
\begin{align}
 &\mathbf{C}_\varepsilon = \Big(\cos\omega \Delta t\,,\; \cfrac{\sin\omega \Delta t}{\omega} \,,\; -\cfrac{\sin\omega \Delta t}{\omega} \,,\; \cfrac{\omega\Delta t\,\cos\omega \Delta t-\sin\omega \Delta t}{\omega^3\Delta t}\,,\;\cfrac{\sin\omega \Delta t-\omega \Delta t}{\omega^3\Delta t}\, \Big)\,,\nonumber\\
  &\mathbf{C}_g = \Big(- \omega\,\sin\omega \Delta t\,,\;\cos\omega \Delta t\,,\; 1-\cos\omega \Delta t \,,\; \cfrac{1-\cos\omega \Delta t-\omega\Delta t\,\sin\omega \Delta t}{\omega^2 \Delta t} \,,\;- \cfrac{1-\cos\omega \Delta t}{\omega^2 \Delta t}\, \Big)\,.\nonumber
\end{align}
\end{widetext}

On any interval, when $\mathbf{j}$ and $n$ can be assumed constant, \cref{maxwell5} is the exact solution of the Maxwell equations. The physical precision of such solution is independent of $\Delta t$, but is defined solely by the accuracies of the initial and boundary conditions. This is a principal advantage of the developed PSATD method along with the high accuracy of spatial derivatives achieved in spectral calculations.

\paragraph{Charge continuity.} In the simulations, \cref{maxwell5} relies on the charge density and current, which are calculated via the weighted projections of macro-particles positions and velocities onto the grid. The numerical errors, produced in such projections, typically dominate the overall accuracy of the model, and should be considered with care. One issue related to the projection errors in PIC, is that the continuity equation,
\begin{equation}\label{continuity_eq}
\partial_t n+\bm{\nabla}\cdot \mathbf{j}=0\,,
\end{equation}
may be not satisfied with the sufficient precision. Consequently, the violation of charge continuity leads to violation of the Poisson equation, $\bm{\nabla}\cdot\mathbf{E}=n$, and the magnetic monopoles absence condition, $\bm{\nabla}\cdot\mathbf{B}=0$. Considering the series of \cref{maxwell5}, one may show that:
\begin{equation}
\widehat{\bm{\nabla}\cdot \bm{\varepsilon}_1} - \widehat{n_1} \propto \widehat{\bm{\nabla}\cdot \bm{g}_1}\propto \frac{\widehat{n_1}-\widehat{n_0}}{\Delta t} +  \widehat{\bm{\nabla}\cdot\mathbf{j}}_{1/2}\,,\nonumber
\end{equation}
where the term in the rightmost side corresponds to the continuity equation \cref{continuity_eq}. The produced errors tend to accumulate, and the unphysical electric and magnetic fields may develop.

Commonly in the PIC methods, the charge continuity is assured by either correcting the deposed current \cite{morse:PoF1971}, or by using the charge-conserving deposition techniques \cite{esirkepov:CPC2001}. The proper current correction may be introduced as, ${\mathbf{j}'=\mathbf{j}-\bm{\nabla}\Gamma}$, where $\Gamma$ satisfies, ${\nabla^2\Gamma = \partial_t n+\bm{\nabla}\cdot \mathbf{j}}$. In the Fourier space, this correction becomes a simple algebraic operation, which makes this approach especially convenient in the spectral methods \cite{Vay:JCP2013}. In the developed scheme, the current correction reads:
\begin{equation}\label{poiss_chk1}
\widehat{\mathbf{j}}_{1/2}' = \widehat{\mathbf{j}}_{1/2} + \cfrac{1}{\omega^2}\left( \cfrac{ \widehat{\bm{\nabla} n}_1 - \widehat{\bm{\nabla} n}_0}{\Delta t} + \widehat{\bm{\nabla}\bm{\nabla}\cdot\mathbf{j}}_{1/2} \right)\,,
\end{equation}
and in contrast to the pure Fourier approach, differential operations in \cref{poiss_chk1} are not linear in the Fourier-Bessel space, but involve matrix operations. In the simulations, this correction should applied at each PIC cycle after the step ({\bf iii}).

\paragraph{Boundary conditions for the fields.} In spectral methods, the boundary conditions of the simulation domain are determined by the basis functions, and providing the model-specific boundaries in the general case can be rather challenging. In our model, the vertical boundaries are periodic, as imposed by the Fourier transform, and the horizontal boundary, $r=R$, is chosen to be reflective (cf \cref{appendix1}). In the situations, when these conditions do not correspond to the physical model, it is often possible to assume the unbounded media, i.e. sufficiently large simulation domain, so that the interaction does not reach the boundaries. In the cases, when such approximation cannot be provided, the absorbing boundaries can be produced by multiplying the concerned values by the evanescent envelopes (for an example, see \cite{kosloff:JCP1986}).

In beam-plasma interactions, the domain of interest may often be restricted to the area around a laser or particle beam, which travels through the plasma. In this case,  we may apply the unbounded media model by considering the simulation domain, which covers the interaction region, and co-propagates with it. Obviously, such \emph{moving window} technique requires the fields at the downstream boundary to be suppressed in order to prevent their upstream translation. In our scheme, when using the moving window, after each shift of the simulation domain we multiply $\bm{\varepsilon}$ and  $\bm{g}$ with a profile function, which is equal to unity everywhere but in a narrow layer near the boundary, where it is evanescent. We have tested the profiles:
$$f(0<x<l_\text{abs}) = \frac{1}{4} \left(1-\cos\frac{\pi x}{l_\text{abs}}\right)^2\,,$$ 
for suppression of $\bm{\varepsilon}$, and
$$f(0<x<l_\text{abs}) = 1 - \mathrm{e}^{-10 x^2/l_\text{abs}^2}\,,$$ 
for a less perturbing suppression of $\bm{g}$. For this operation, the fields are projected to the real space along $x$-coordinate via the inverse Fourier transform, and returned to the spectral domain after the profiling. Plasma is also removed from this layer. The physical properties of such "absorbing" layers are also affected by the fact, that they travel with the domain. In the performed tests, the described method has proven efficient, however, its more rigorous development remains a subject for the further studies.

\section{\label{sec2} Simulations}

The described algorithm was implemented as the separate module in the code PLARES \cite{andriyash:JCP2014}. The original code was designed to simulate physics of free electron lasers using the reduced Fourier and Fourier-Bessel Maxwell solvers acting directly on the particles. In PLARES-PIC we use the linear interpolation for the particle-grid projections, and the particles are advanced in the $(\mathbf{r},\mathbf{p})$-space via the standard Boris pusher \cite{boris:proc1973}.

Code runtime is managed by the scripts written in Python, which provides very simple coding and takes benefit of the numerical and scientific computation modules, Numpy and Scipy \cite{numpy_ref}, and on-the-fly simulation visualization can be easily implemented using the Matplotlib module \cite{pylab_ref}. To provide the higher code performance, the computationally intense operations are written in Fortran 90, and are wrapped for Python calls via F2PY interface generator \cite{f2py_ref}. For the fast Fourier transforms we use the FFTW3 package \cite{fftw3_ref}. The parallel computation is managed by MPI from the Python runtime via MPI4PY \cite{mpi4py_ref}. For simplicity in our code we use the radial decomposition, i.e. the spatial grid and the particles are divided into the slices along the radial direction. 

\subsection{\label{sec2:1} Linear laser-plasma interaction}

Let us firstly check of the scheme dispersion properties, by modeling the propagation of laser pulse in vacuum and plasma. The classical dispersion relation of electromagnetic waves in plasma reads, ${\omega^2=k^2c^2+\omega_{pe}^2}$, where $\omega_{pe}=(4\pi e^2 n_e/m_e)^{1/2}$ is the frequency of electron plasma with the density $n_e$ (is cgs units). In the underdense plasmas, where $\omega_{pe}\ll \omega_0$, a simple estimate for the radiation group velocity can be derived, ${\beta_G = \partial_{kc}\omega \simeq 1-\omega_{pe}^2/2k^2c^2}$. Moreover, even in vacuum the finite-size laser beam is slowed by the diffraction \cite{esarey:JOSAB95}. As a result, the deviation of the Gaussian beams centroid velocity from the speed of light in vacuum estimates as:
\begin{equation}\label{group_veloc}
 1-\beta_G = n_e/2\,n_c + (\lambda_0/2\pi w_0)^2\,,
\end{equation}
where $n_e$ is the plasma density, and $w_0$ is the beam waist. 

\begin{figure}\centering
 \includegraphics[width=0.49\textwidth]{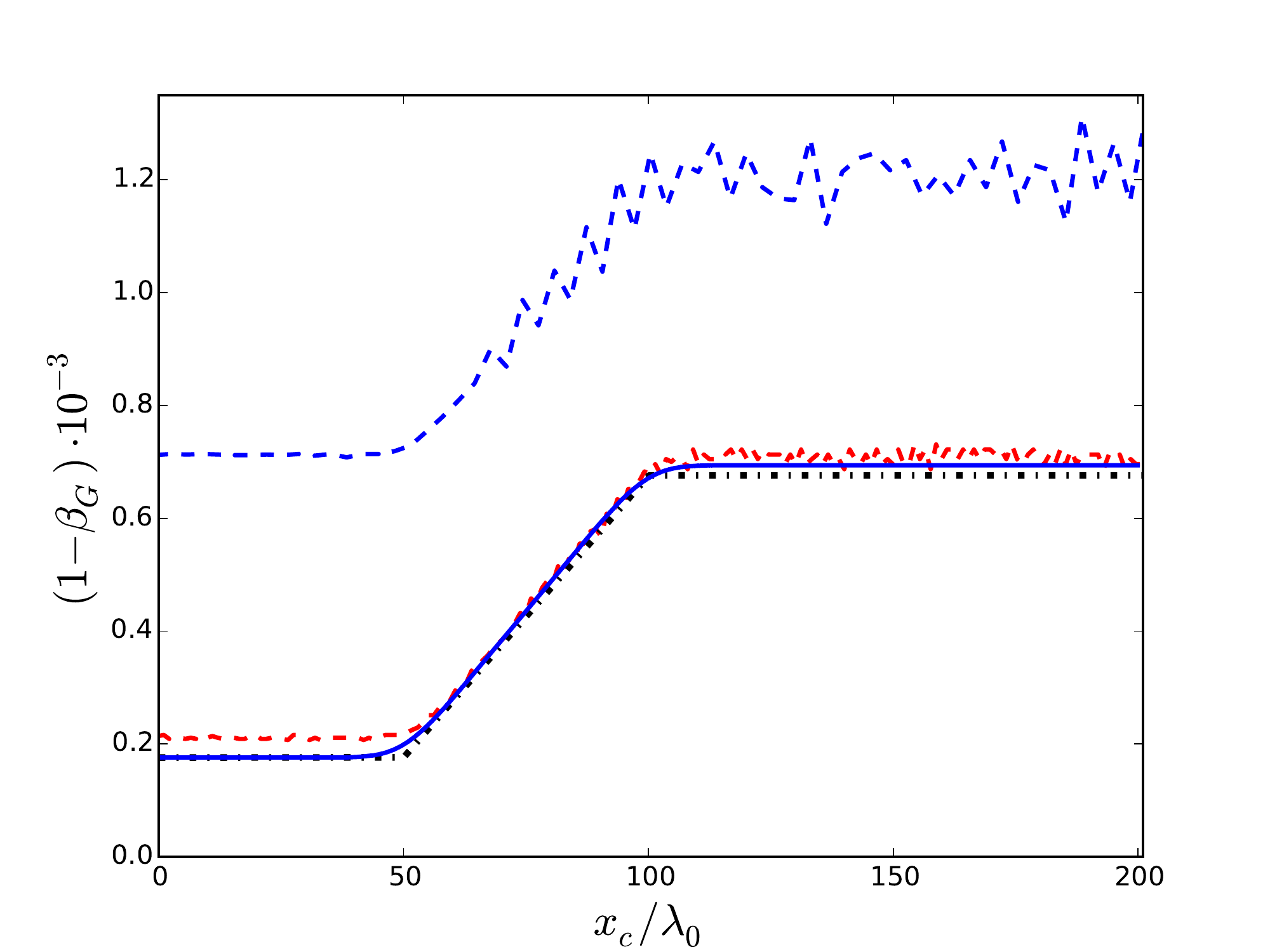} 
\caption{Group velocity of the laser beam in the PSATD PIC simulation with $\Delta x=0.048\,\lambda_0$ (solid blue curve), and in FDTD PIC simulations with $\Delta x=0.048\,\lambda_0$ (dashed blue curve), and $\Delta x=0.016\,\lambda_0$ (dashed red curve). The  black dot-dashed line corresponds to the theoretical estimate \cref{group_veloc}.}\label{fig1}
\end{figure}

In our test, we propagate the linearly polarized laser beam with the Gaussian profiles, $a = a_0\exp(-r^2/w_0^2-x^2/l_x^2)$, where the parameters are $a_0 = 10^{-2}$, $w_0 = 12 \lambda_0$, and $l_x = 12 \lambda_0$. Firstly, laser travels $50\lambda_0$ distance in vacuum until its centroid enters the plasma. The plasma density increases linearly along first $50\lambda_0$, and then reaches its maximal value of $n_e = 10^{-3} n_c$, after which it remains constant. The longitudinal and transverse resolutions are $\Delta x=0.048\lambda_0$ and  $\Delta r=0.32\lambda_0$, and the time-step is $\Delta t = \Delta x/c$. At each iteration we measure the laser beam centroid  position as $x_c = \sum x r E_z^2/\sum r E_z^2$, and then deduce its group velocity $\beta_G=\Delta x_c/\Delta t$, which is also averaged  over the laser period.

Evolution of the laser group velocity in our simulation is shown in \cref{fig1} with a blue solid curve, and theoretical estimate \cref{group_veloc} is plotted with the black dot-dashed line. The blue dashed line corresponds to the FDTD PIC simulation, performed for the same grid resolutions with the quasi-cylindrical code CALDER-CIRC \cite{lifschitz:JCP2009}. One can see that the numerical dispersion in FDTD PIC,  significantly slows the laser free-space propagation and perturbs its propagation in plasma. To approach the correct laser velocity with the finite differences method, we had to use much higher resolutions, $\Delta x = 0.016 \lambda_0$, $\Delta r=0.16 \lambda_0$ (red dashed curve).

Propagating in plasma, the laser excites the plasma wave, often refereed as a \emph{wake}. When $a_L^2\ll1$, the produced wake is "linear", i.e. the density perturbations in electron plasma are very small, if compared to the plasma density. The laser-driven plasma waves have been extensively studied, and in the linear regime the generated electrostatic fields, aka wakefields, can be described analytically (cf \cite{gorbunov:JETP1987}). From this linear theory, the longitudinal wakefield reads:
\begin{equation}\label{lin_wake}
  \varepsilon_x = \left(\frac{\omega_{pe}}{2 \omega_0}\right)^2\int_{x}^{\infty}\tilde{\varepsilon}_L^2 \cos[k_{pe}(x-x')]\mathrm{d}x'\,,
\end{equation}
where the electron plasma wavenumber is $k_{pe}=\omega_{pe}/c$.

\begin{figure}\centering
 \includegraphics[width=0.49\textwidth]{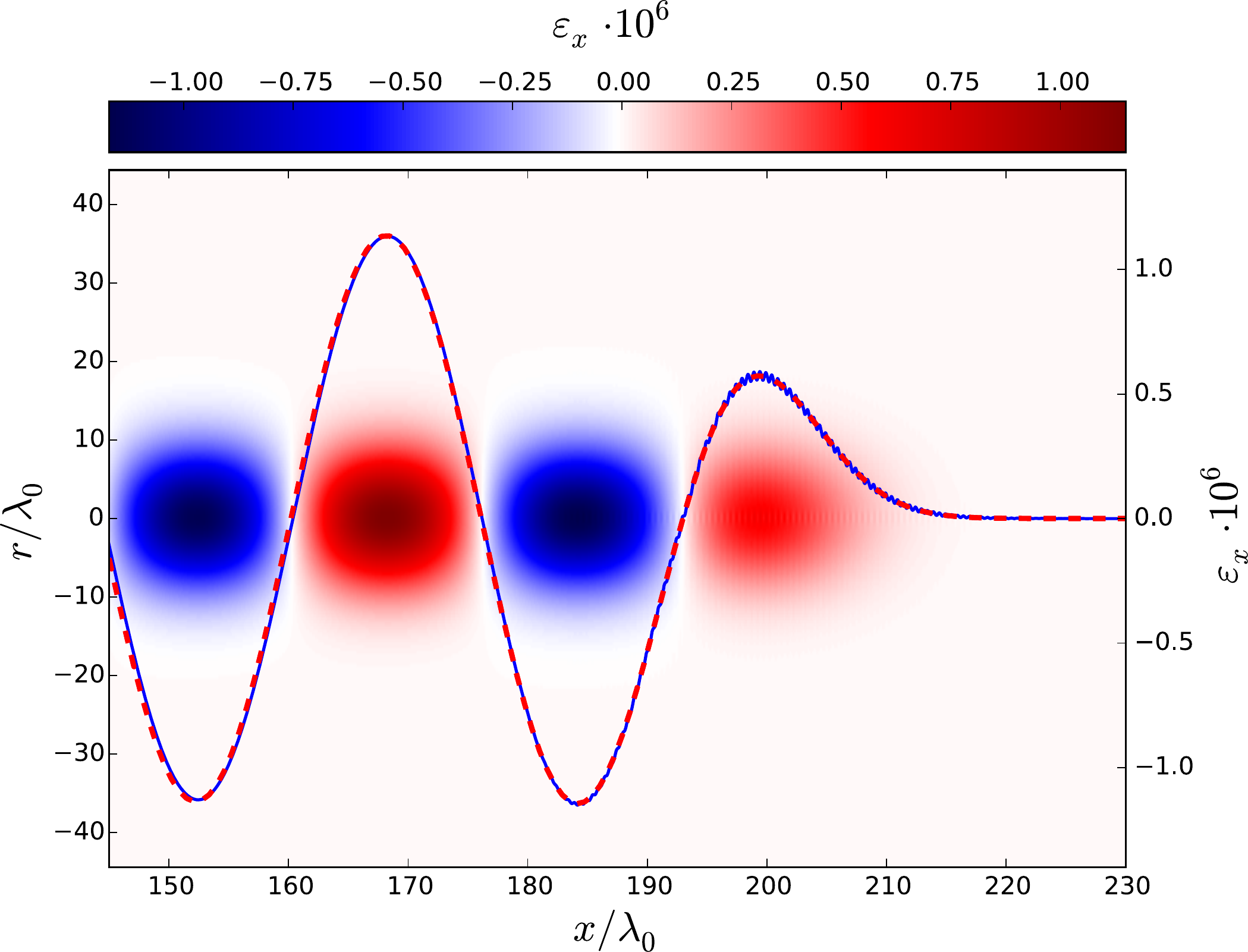} 
\caption{Map of the longitudinal electric field $\varepsilon_x$ in plasma. Distribution of the on-axis field value $\varepsilon_x(r=0)$ extracted from the simulation (solid blue curve) and predicted by the theory (dashed red curve);}\label{fig2}
\end{figure}

In \cref{fig2}, we map with colors the distribution of electrostatic field $\varepsilon_x$, which clearly corresponds to the plasma wave. The solid blue curve shows the field at the axis $\varepsilon_x(x,r=0)$, and it is compared to the estimate provided by \cref{lin_wake} shown by the dashed curve.

\subsection{\label{sec2:2} Laser plasma acceleration of electrons}

\begin{figure*}\centering
\includegraphics[width=0.9\textwidth]{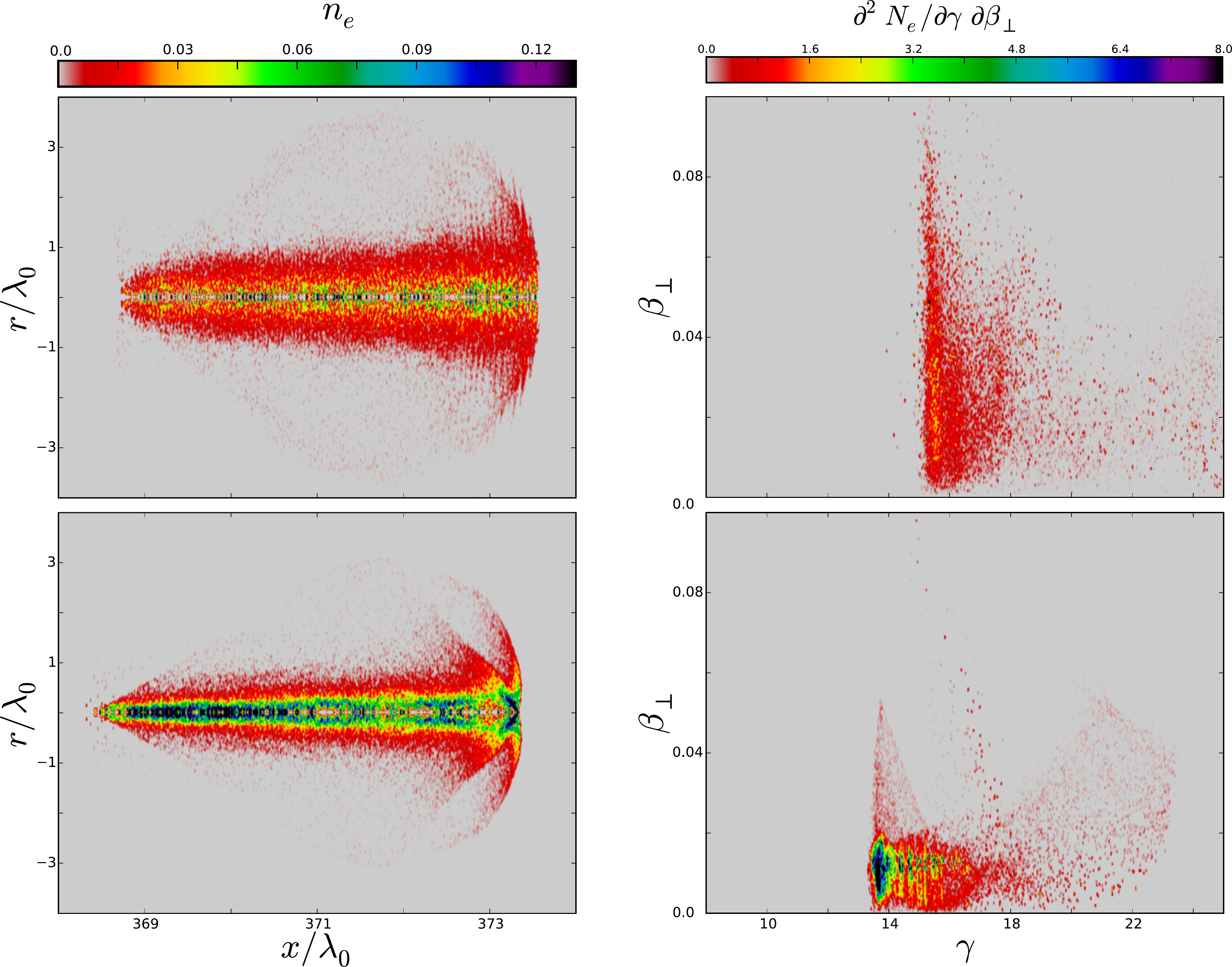}
\caption{Maps of densities (left panel) and spectra (right panel) of accelerated electrons  modeled with FDTD (upper figures) and PSATD (lower figures) methods.}\label{fig3}
\end{figure*}

For a more complex test, we model the acceleration of electrons from the underdense plasma by a few-mJ-few-cycle laser pulse. The practical interest to this mechanism is related to the recent idea of using the high repetition rate (kHz) laser for the sources of low energy femtosecond electron beams \cite{beaurepaire:NJP2014}. Tightly focusing such a laser in a sub-millimeter short plasma, one produces a strong bubble-like wake along the laser Rayleigh length, which is only a few dozens of micrometers. To provide the electron injection into this wake, an abrupt change of the plasma density, so-called \emph{shock}, is produced near the laser focus. Such interaction is rich with laser-plasma physics e.g.  the laser-driven wakefield, which evolves from the linear to the bubble regime,  laser self-focusing near the shock, and generation of the secondary wake by the accelerated beam. 

In our test, we consider the pre-ionized plasma with the density profile, which grows linearly along first $300\lambda_0$ to reach its maximal value of $n_{max} = 0.005\,n_c$, then rapidly falls to $n_e =0.5\,n_{max}$ over 15 $\lambda_0$, and further remains constant. The Gaussian laser beam with $a_0 = 3$, $w_0 = 4 \lambda_0$, and $l_x = 5 \lambda_0$ is focused at the density peak at $x=300\lambda_0$. The  interaction is visually demonstrated in \cref{supp}, where the animation is provided by the graphical output from the PLARES-PIC simulation. 

For a more detailed analysis, we have compared this simulation with the same test performed with CALDER-CIRC code. Note, that the group velocity of such a tightly focused laser differs significantly from the speed of light in vacuum, hence, the simulation is rather sensitive to the numerical dispersion. In PSATD simulation we consider the grid with $\Delta x=0.025\lambda_0$ and  $\Delta r=0.25\lambda_0$, and the time step is $\Delta t= \Delta x/c$. To resolve correctly the laser propagation, and hence the electron injection, the FDTD simulation requires higher resolution, and we used $\Delta x=0.016\lambda_0$ and  $\Delta r=0.16\lambda_0$, and the time step was $\Delta t= 0.98\,\Delta x/c$. 

In both simulations the structures of the plasma wakes are almost identical, while the accelerated beams differ significantly. To study this we select the electrons with $\gamma_p>8$ (injected ones), and project their density onto the $(x,r)$ and $(\gamma,\beta_\perp)$ planes for the time $385 \lambda_0/c$ after laser enters the plasma. In the right of \cref{fig3} we plot the $(x,r)$  density profiles, and, while in PSATD  beam has a clear structure with the injection signatures,  in FDTD simulation the beam is blurred, and we see the small-scale density modulations near its front. These result from the artificial high-frequency electromagnetic waves, which in FDTD are significantly slowed by numerical diffraction, and are generated by the relativistic particles, in a way similar to the physical Cherenkov radiation in the dispersive media. The numerical Cherenkov effect, along with the errors of Lorentz force projection, are known to also increase beams emittance in the wakefield acceleration simulations \cite{lehe:PRSTAB2013,lehe:PRSTAB2014}. The particles transverse velocities FDTD  is significantly affected by the numerical effects (see upper left of \cref{fig3}), and for more of comparative analysis of the numerical effects in the quasi cylindrical FDTD and PSATD methods see \cite{lehe:arxiv2015}. 

The electrons energy in FDTD simulation is slightly bigger than one in PSATD, as well as the full charge, which is 33 pC and 28 pC in the FDTD and spectral models respectively  (for $\lambda_0=0.8\mu$m). These differences can be attributed to the electron injection modeling, which in this test is very fast and is therefore very sensitive to the laser velocity.

\section{\label{sec3} Conclusions}

We have developed and discussed a spectral quasi-cylindrical particle-in-cell method designed for plasma modeling. The proposed scheme is based on the pseudo-spectral analytical time domain method, and, in contrast to the commonly used finite-difference PIC schemes, it does not produce grid-related numerical dispersion of electromagnetic fields. Moreover, the electric and magnetic fields are not staggered, which significantly reduces the errors of the Lorentz force projection. This makes the proposed approach advantageous in a wide class of problems, where co-propagation of relativistic particles and light is involved, e.g. simulations of the wakefield accelerators.

The developed model was implemented in the code PLARES-PIC, which was tested and benchmarked against the quasi-cylindrical FDTD PIC code CALDER-CIRC. The new code has demonstrated capacity to accurately model the laser propagation in vacuum and in plasma, even with the spatial and temporal resolutions few times lower than in FDTD PIC. In the more complex test, we have simulated wakefield acceleration of electrons by a few-mJ-few-cycle laser in the configuration, where electrons are injected in the plasma density shock. The new spectral code has demonstrated a good agreement with the FDTD PIC, save for the numerical effects development in the latter.

The proposed method provides a principally new approach to three-dimensional modeling of plasma electrodynamics. Although the spectral transforms are computationally more demanding than conventional FDTD PIC, the natural accuracy of the spectral method and absence of the numerical dispersion, can often compensate the additional load by using lower temporal and spectral resolutions.

\section*{Acknowledgments}

One of the authors (IAA) acknowledges the support of Victor Malka and Marie-Emmanuelle Couprie, and partial funding from their ERC programs X-Five (contract No.~339128) and COXINEL (contract No.~340015).

\section*{\label{supp} Supplementary Materials}

In \cite{plaresmovie}, we use green colors to plot the electron density, and the colormap scales from the dark green at $n_e=0$ to the white at $n_e=0.01\,n_c$. The amplitude of the laser beam is mapped in semi-transparent red colors, and their colormap maximum is fixed at $\varepsilon_z = 3$.  The video is based on 665 images, which the graphical module of PLARES-PIC outputs each 30 time-steps.

\appendix
\section{\label{appendix1} Definition of Fourier-Bessel transform}

The generalized Fourier-Bessel series consists of the \textit{cylindrical harmonics}, ${\mathcal{H} = \mathrm{e}^{i k_x x + i m \theta}J_m(k_r r)}$, where $(x,r,\theta)$ are the coordinates, $J_m$ is the $m$-th order Bessel function, and parameters $(k_x,k_r,m)$ are the coordinates in the spectral space. In the spectral transformation these harmonics are used to represent a function defined in a three-dimensional space as:
\begin{equation}\label{decompose}
f(\mathbf{r}) = \sum_{\mathbf{k}\in \mathcal{K} }\mathcal{H}(\mathbf{k},\mathbf{r}) \,\widehat{f}_\mathbf{k} \,,
\end{equation}
where the discrete spectral domain $\mathcal{K}$ defines the possible wavenumbers $\mathbf{k}=(k_x\,,k_r\,,m)$. Practically, \cref{decompose} corresponds to the three transforms of the initial function: Fourier transform over the angle $\theta$, Fourier transform over $x$-axis, and Hankel transform in the radial direction. 

In the quasi-cylindrical PIC methods, the angular Fourier decomposition is considered at the particles positions \cite{lifschitz:JCP2009}. This means, that the weight of each macro-particle is presented as $w_p=\sum_m w_p^{(m)} \mathrm{e}^{-i m \theta_p}$, and the components $w_p^{(m)}$ are gathered on the grid separately, to obtain the azimuthal components of the charge density and current. When the angular components of the fields $\mathbf{E}_p^{(m)}$ are calculated from electromagnetic equations, the total fields are projected onto the particles as, $\mathbf{E}_p = \sum_m \mathrm{e}^{i m \theta_p} \mathbf{E}_p^{(m)}$ . 

The spectral decomposition in $(x,r)$-space is done at the nodes of the grid via the consecutive discrete Fourier and Hankel transforms (DFT and DHT). For the DFT we use the fast Fourier transform (FFT), which operates on the uniform spatial and spectral grids, and involves only $\sim N_x \log N_x$ instead of the standard $N_x^2$-fold DFT matrix product. Note, that FFT naturally provides the periodic boundaries at $x=0$ and $x=N_x\Delta x$.

For the radial transforms, we define the matrices of the inverse Hankel transforms as, ${\text{IDHT} = J_m(k_r^{(m)} r_j)}$, and consider the uniform spatial grid $r_j = j R/N_r$, where $j=1,..,N_r$ are integers, and $R$ is a radial boundary of the simulation domain. Note, that the radial grid is common for all modes, and the point $r=0$ is excluded, so that the matrices for $m>0$ will not be degenerate. In this case, the DHT matrix can be computed via the numerical inversion $\mathrm{DHT} = \mathrm{IDHT}^{-1}$. 

The choice of the spectral grids $k_r^{(m)}$ defines the properties of the radial boundaries for each azimuthal mode. We choose the radial boundary $r=R$, to satisfy the Dirichlet condition, $f(R)\equiv 0$, for which $k_r^{(m)} = u_j^{(m)}/R$, where $u_j^{(m)}$ are the zeros of $J_m$. In the continuous space, such choice provides the orthogonality of the Bessel terms, and assures the transforms reversibility.

\section{\label{appendix2} Useful differential properties of Fourier-Bessel transform}

Mathematical properties of the Fourier-Bessel series are well known, and here we briefly revise their main features used in our study. The differential properties of  \cref{decompose} are defined by the basis functions $\mathrm{e}^{i k_x x}$, $\mathrm{e}^{i m \theta}$ and $J_m(k_r r)$. It is easy to see that, cylindrical harmonics are the eigenfunctions of the Laplace operator, i.e. $\widehat{\nabla^2 f} = -\omega^2\widehat{f}$, where $\omega = \sqrt{k_x^2+ k_r^2}$.

Let us, construct the first-order differential operators for the Cartesian vector components. For this, we note that the derivative over the longitudinal coordinate is $\partial_x=i k_x$, and the transverse derivatives can be presented as, 
$$\partial_y = \cos\theta\, \partial_r -\frac{\sin\theta}{r}\, \partial_\theta\,,\quad \partial_z = \sin\theta\, \partial_r-\frac{\cos\theta}{r}\,\partial_\theta\,.$$ 
Using the properties of the Bessel functions, we calculate the components of the scalar function gradient as:
\begin{align}\label{diff_fb}
&\partial_x f^{(m)} = \text{IDFT}_{x\,k_x} * \text{IDHT}^{(m)}_{r\,k_r} * i\, k_x\,\widehat{f}_{k_x\,k_r}^{(m)}\,,\nonumber\\
&\partial_y f^{(m)} = \text{IDFT}_{x\,k_x}* \Big(\partial_\perp\text{IDHT}^{(m+1)}_{r\,k_r} * \widehat{f}_{k_x\,k_r}^{(m+1)} - \nonumber\\
& \qquad-\partial_\perp\text{IDHT}^{(m-1)}_{r\,k_r} * \widehat{f}_{k_x\,k_r}^{(m-1)} \Big)\,, \\
&\partial_z f^{(m)} = \text{IDFT}_{x\,k_x}* \Big(\partial_\perp\text{IDHT}^{(m+1)}_{r\,k_r} * i\,\widehat{f}_{k_x\,k_r}^{(m+1)} + \nonumber\\
& \qquad+\partial_\perp\text{IDHT}^{(m-1)}_{r\,k_r} * i\,\widehat{f}_{k_x\,k_r}^{(m-1)}\Big) \,, \nonumber
\end{align}
where "$*$" means matrix product, and the transformation matrices for the transverse derivatives are:
$$ \partial_\perp\text{IDHT}^{(m\pm1)}_{r\,k_r} = \cfrac{k_r^{(m\pm1)}}{2}\, J_m\left(k_r^{(m\pm1)} r\right)\,.$$
In contrast to the ordinary Fourier series, the transverse derivatives in \cref{diff_fb} couple the azimuthal modes $m$ with $m+1$ and $m-1$.

Linearly combining the transformation matrices, one can construct any necessary differential operatior. For example, for the spectral-spectral and real-spectral derivative projectors one should use the operators, $$\text{DHT}^{(m)}_{k_r\,r}*\partial_\perp\text{IDHT}^{(m\pm1)}_{r\,k_r}\,,$$ and, $$\text{DHT}^{(m)}_{k_r\,r}*\partial_\perp\text{IDHT}^{(m\pm1)}_{r\,k_r}*\text{DHT}^{(m\pm1)}_{k_r\,r}\,,$$ respectively. 

In our study, we use \cref{diff_fb} for spectral-real projection of the magnetic field, and another differential operator is constructed for the real-spectral projection of charge density $n$ directly to its gradient $\widehat{\bm{\nabla}n}$. The spectral-spectral \textbf{rot}, \textbf{div} and \textbf{grad} operations are used for the current correction operations in \cref{poiss_chk1} and for magnetic field calculation \cref{maxwell2:2}.

\end{document}